\documentstyle[aps,preprint,epsfig,floats]{revtex}
\begin{document}
\tightenlines
%% the segment below is from prabib.sty, and has been altered so that a new
%% page is not started for the references (saves paper)
\catcode`@=11
\def\references{%
\ifpreprintsty
%\newpage
\bigskip\bigskip
\hbox to\hsize{\hss\large \refname\hss}%
\else
\vskip24pt
\hrule width\hsize\relax
\vskip 1.6cm
\fi
\list{\@biblabel{\arabic{enumiv}}}%
{\labelwidth\WidestRefLabelThusFar  \labelsep4pt %
\leftmargin\labelwidth %
\advance\leftmargin\labelsep %
\ifdim\baselinestretch pt>1 pt %
\parsep  4pt\relax %
\else %
\parsep  0pt\relax %
\fi
\itemsep\parsep %
\usecounter{enumiv}%
\let\p@enumiv\@empty
\def\theenumiv{\arabic{enumiv}}%
}%
\let\newblock\relax %
\sloppy\clubpenalty4000\widowpenalty4000
\sfcode`\.=1000\relax
\ifpreprintsty\else\small\fi
}
\catcode`@=12

\hfuzz=4pt

\preprint{\font\fortssbx=cmssbx10 scaled \magstep2
\hbox to \hsize{
\includegraphics{uwlogo.ps}
\hskip.5in \raise.1in\hbox{\fortssbx University of Wisconsin - Madison}
\hfill$\vcenter{\tighten
\hbox{\bf MADPH-97-1022}
\hbox{\bf hep-ph/9711378}
\hbox{November 1997}}$}}

\title{\vspace*{.35in}  %% add a little space before title
On the LEP Measurements of the Rising Photon-Photon\\ Total Cross Section}

\author{M.\ M.\ Block}
\address{Department of Physics, Northwestern University \\
Evanston, IL 60208}

\author{E.\ M.\ Gregores and F.\ Halzen}
\address{Department of Physics, University of Wisconsin \\
Madison, WI 53706}

\author{G.\ Pancheri}
\address{INFN - Laboratori Nazionali di Frascati \\
Frascati, Italy}

\maketitle
\thispagestyle{empty}

\begin{abstract}
\vskip-5ex %% reduce space between "Abstract" and it's text

Using vector meson dominance we calculate the $\gamma\gamma$ total cross  
section from data on $pp$, $p\bar p$ and $\gamma p$ total cross sections. Our  
result agrees with recent high energy measurements performed by the L3  
experiment at the LEP collider, not with the conflicting preliminary
data obtained by OPAL.

\end{abstract}
\draft
\pacs{ }

%%%%%%%%%%%%%%%%%%%%%%%%%%%%%%%%%%%%%%%%%%%%%%%%%%%%%%%%%%%%%%%%%%%%%%

Two experiments at LEP have measured the high energy $\gamma\gamma$
total cross section\cite{exp}. While these measurements yield new
information on its high energy behavior at center-of-mass energies in
excess of $\sqrt{s}=15$ GeV, they may represent the last opportunity
to measure the $\gamma\gamma$ cross section, and the two data sets
unfortunately disagree. We here point out that a simultaneous analysis
of $pp$, $p\bar p$, $\gamma p$ and $\gamma\gamma$ total cross sections
in the framework of vector dominance (VMD) nicely accommodates the L3
measurements. The analysis is sufficiently restrictive to
exclude the preliminary OPAL results.

We have performed a simultaneous analysis of all data on forward
scattering amplitudes involving hadrons and photons in the context of
a model where high energy cross sections rise as a consequence of the
increasing numbers of soft quarks and gluons populating the colliding
particles\cite{margolis}. This idea is implemented in an eikonal
formalism where the total cross section is given by

\begin{equation}
\sigma_{\mbox{\scriptsize{tot}}}(s) =
2P_{\mbox{\scriptsize{had}}}\int\left[1-e^{-\chi_{\rm I}}(s,b)\cos(\chi_{\rm R}(s,b)\right]\, d^2\vec{b}\, ,
\end{equation}
with the eikonal $\chi$ given by $\chi(s,b)=\chi_{\rm R}(s,b)+i\chi_{\rm I}(s,b)$.
Here $b$ is the impact parameter and $P_{\mbox{\scriptsize{had}}}$ the
probability that the particle interacts as a hadron. Its value is
obviously one for the proton and antiproton, and of order $\alpha$
when the particle is a photon. Its value can be obtained from VMD. We
will use the value $P_{\mbox{\scriptsize{had}}}=1/240$ for $\gamma p$
collisions and, therefore, $P_{\mbox{\scriptsize{had}}}=(1/240)^2$ for
$\gamma\gamma$ collisions.

The eikonal $\chi$ receives contributions from quark-quark,
quark-gluon and gluon-gluon interactions, therefore
\begin{eqnarray}
\chi(s,b) &=& \chi_{qq}(s,b)+\chi_{qg}(s,b)+\chi_{gg}(s,b)
\nonumber \\ 
&=& A(\mu_{qq},b)\sigma_{qq}(s)
+ A(\sqrt{\mu_{qq}\mu_{gg}},b)\sigma_{qg}(s)
+ A(\mu_{gg},b)\sigma_{gg}(s) \, ,
\end{eqnarray}
where $A(\mu,b)$ is the overlap function in impact parameter space and  
$\sigma_{ij}$ the cross sections of the colliding partons. We simply  
parametrize $A(\mu,b)$ as the Fourier transform of a dipole form factor
\begin{equation}
A(\mu,b)=\frac{\mu^2}{96\pi}(\mu b)^3 K_3(\mu b)\, ,
\end{equation}
where $K_3(x)$ is the modified Bessel function of second kind. Since the
``size'' of quarks and gluons in the proton can be different, we
obtain $\mu_{qq}=0.89$ and $\mu_{gg}=0.73$. This model where hadrons
asymptotically evolve into black disks of partons nicely accommodates
all data on $pp$ and $p\bar p$ total cross sections as well as
measurements of the differential elastic cross section and of the
real-to-imaginary part of the forward scattering amplitude\cite{future}.
The results shown in Fig.~1 have been updated for a new measurement of the
$p\bar p$ total cross section at the Fermilab collider\cite{newpp}.

Following reference\cite{fletcher} we obtain $\gamma p$ and
$\gamma\gamma$ total cross sections from the assumption that, in the
spirit of VMD, the photon is a 2 quark state in contrast with the
proton which is a 3 quark state. The $\gamma p$ total cross section is
obtained from the even eikonal (used in the $pp$ and $p\bar p$ fit of Fig.~1) by the substitutions:
\[\begin{array}{l}
\sigma_{ij} \rightarrow \frac{2}{3} \,\sigma_{ij} \\
\mu_i \rightarrow \sqrt{\frac{3}{2}} \,\mu_i 
\, .
\end{array}\]
To obtain the $\gamma\gamma$ total cross section, we  now substitute
\[\begin{array}{l}
\sigma_{ij} \rightarrow \frac{4}{9} \,\sigma_{ij} \\
\mu_i \rightarrow \frac{3}{2}\,\mu_i \, 
\end{array}\]
into the nucleon-nucleon even eikonal.

The results are shown in Figs.~2, 3. The agreement with data is
impressive. The formalism also accommodates the HERA measurements of
elastic $\rho$ photoproduction. It is clear that VMD selects the L3
and rejects the preliminary OPAL results\cite{exp}.

We have performed this analysis using the model of
reference\cite{margolis}, and repeated it by performing a calculation of
the parton cross section in the mini-jet model\cite{minijet}. The
conclusions are identical and the fits indistinguishable. Details will
be published elsewhere\cite{future}.

\acknowledgments

One of us (MMB) was  partially supported by Department of Energy grant 
DA-AC02-76-Er02289 Task B. This research
was also supported in part by the University of Wisconsin Research
Committee with funds granted by the Wisconsin Alumni Research
Foundation, by the U.S.\ Department of Energy under Grant
No.~DE-FG02-95ER40896, and by Funda\c{c}\~ao de Amparo \`a Pesquisa do
Estado de S\~ao Paulo (FAPESP).

%%%%%%%% REFERENCES %%%%%%%%%%%%%%%%%%%%%

%%%%%%%% FIGURES %%%%%%%%%%%%%%%%%%%%%
\newpage
%****************
\begin{figure}
\begin{center}
\mbox{\epsfig{file=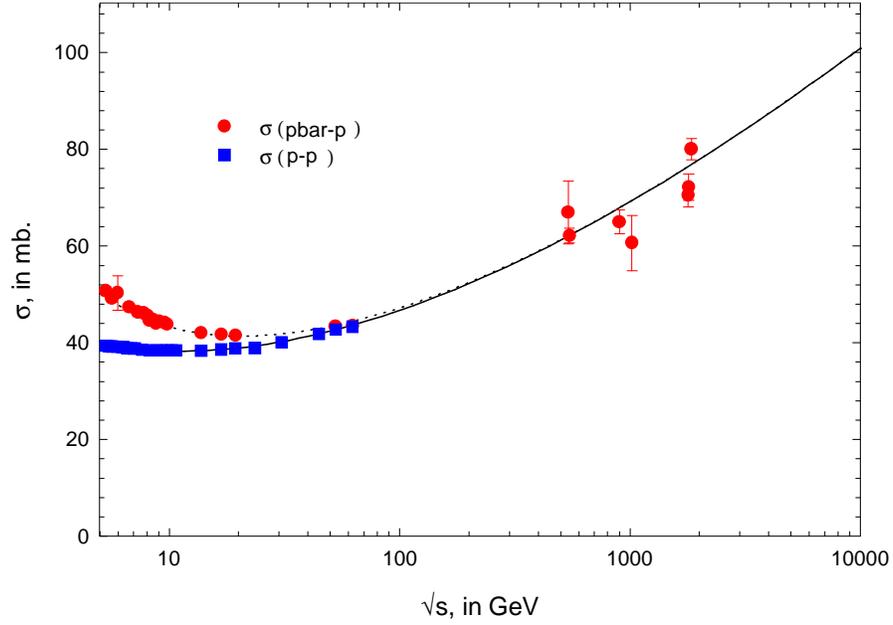%
              ,width=5in,%
bbllx=96pt,bblly=340pt,bburx=550pt,bbury=650pt,clip=}}
\end{center}
\caption[]{The total cross section for $pp$ and $p\bar p$ scattering 
in mb, $\sigma$ {\em  vs} $\sqrt s$, the center-of-mass energy in GeV. 
The solid line and 
squares are for $pp$ scattering and the dotted line and circles are for
 $p\bar p$ scattering.}
\label{fig:pp}
\end{figure}

%****************
\begin{figure}
\begin{center}
\mbox{\epsfig{file=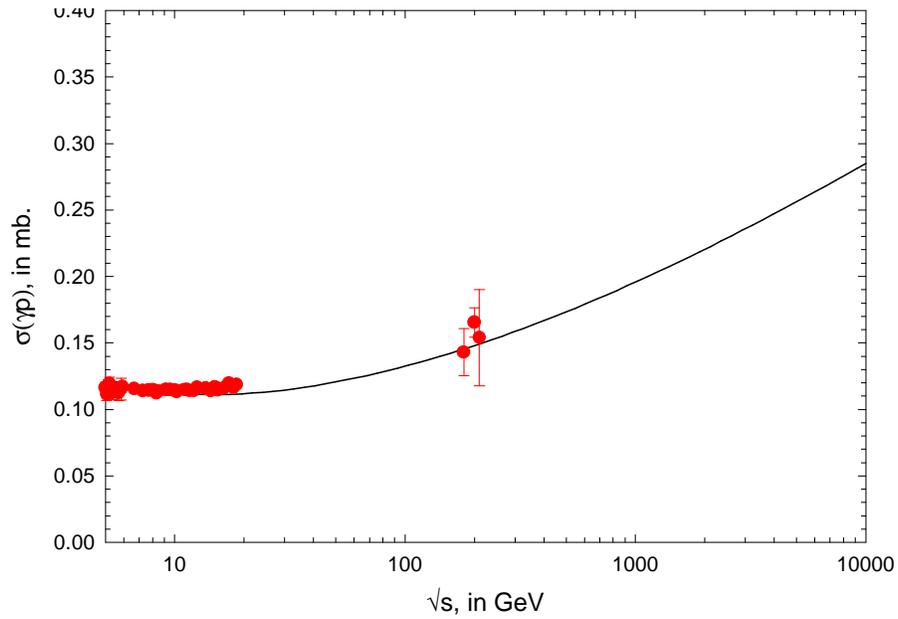%
,width=5in,bbllx=96pt,bblly=340pt,bburx=550pt,bbury=650pt,clip=%
              }}
\end{center}
\caption[]{The total $\gamma p$ cross section in mb, $\sigma (\gamma p)$ 
{\em vs} $\sqrt s$, the center-of-mass energy  in GeV. }
\label{fig:gp}
\end{figure}

%****************
\begin{figure}
\begin{center}
\mbox{\epsfig{file=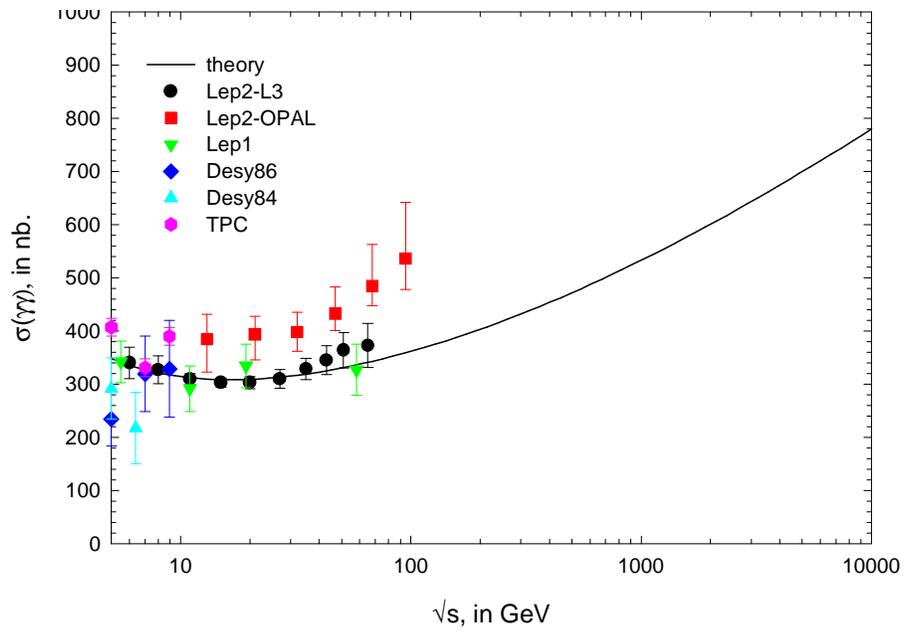%
              ,width=5in,bbllx=96pt,bblly=340pt,bburx=550pt,bbury=650pt,clip=%
}}
\end{center}
\caption[]{The total $\gamma\gamma$ cross section in nb, $\sigma (\gamma\gamma)$
{\em vs} $\sqrt s$, the center-of-mass energy in GeV. }
\label{fig:gg}
\end{figure}

%****************
\end{document}